\documentclass{aa} 

\usepackage{graphicx,psfig}

\newcommand{\Lsun} {L$_\odot$}

\newcommand{\simless}{\mathbin{\lower 3pt\hbox 
      {$\rlap{\raise 5pt\hbox{$\char'074$}}\mathchar"7218$}}} 
\newcommand{\simgreat}{\mathbin{\lower 3pt\hbox 
     {$\rlap{\raise 5pt\hbox{$\char'076$}}\mathchar"7218$}}}  
 
\begin{document}

\title{The Shape of the Inner Rim in Proto-Planetary Disks}

\author{
Andrea Isella \inst{1,2} 
 and  
Antonella  Natta  \inst{1} 
} 
\institute{ 
  Osservatorio Astrofisico di Arcetri, INAF, Largo E.Fermi 5, 
  I-50125 Firenze, Italy 
  \and 
  Dipartimento di Fisica, Universit\'a di Milano, Via Celoria 16,
  20133 Milano, Italy  
} 
 
\offprints{isella@arcetri.astro.it} 
\date{Received ...; accepted ...} 
 
\authorrunning{ISELLA \& NATTA} 
\titlerunning{INNER DISK}

\abstract{  
This paper  discusses the properties  of the inner puffed-up rim which
forms in circumstellar disks when dust evaporates. We argue that the
rim shape is controlled by a fundamental property of circumstellar
disks, namely their very large vertical density gradient, through the
dependence of grain evaporation temperature on gas density. As a
result, the bright side of the rim is {\it curved}, rather
than {\it vertical}, as expected when a constant evaporation
temperature is assumed. We have computed a number of rim models, which
take into account this effect in a self-consistent way. The results
show that  the curved rim (as the vertical rim) emits most of its
radiation in the near and mid-IR, and provides a simple explanation to
the observed values of the near-IR excess (the ``3 $\mu$m bump" of
Herbig Ae stars). Contrary to the vertical rim, for curved rims the
near-IR excess does not depend much on the inclination, being maximum
for face-on objects.
We have then computed synthetic images of the curved rim seen
under different inclinations; 
face-on rims are seen as bright, centrally symmetric rings
on the sky; increasing the inclination, the rim takes an elliptical
shape, with one side brighter than the other.
}
\maketitle

\section {Introduction} 
 
The structure of the inner regions of  circumstellar disks associated  
with pre-main sequence stars is the subject of intense research.  
Interferometers working in the near infrared are providing the first  
direct information on the morphology of disks on scales of fractions  
of AU. They show that in the majority of cases the observed visibility  
curves are not well reproduced by flared disk models; rather, they 
are consistent with the emission of a ring of uniform brightness,  
of radius similar to the dust evaporation distance from the star 
(Millan-Gabet et al. \cite{M01}, Tuthill et al. \cite{T01}).  
 
The interferometric results provide strong support to the idea that 
the inner disk structure deviates substantially from that of a flared 
disk because dust evaporation introduces a strong discontinuity in the 
opacity, which results in a ``puffed-up" rim at the dust destruction 
radius, where dust is exposed directly to the heating stellar 
radiation. The idea of a puffed-up inner rim was proposed by Natta et 
al. (\cite{N01}) and developed further by Dullemond et
al. (\cite{DDN01}, hereafter DDN01) for Herbig Ae stars, to account
for the shape of the near-infrared excess of these stars (the
``3-$\mu$m bump"). These authors  pointed out that the rim had
also the right properties to explain the early interferometric
results of Millan-Gabet et al. (\cite{M01}). Recent theoretical work
by Muzerolle et al. (\cite{MAC04}) has shown that the condition
required to produce a puffed-up inner rim are indeed likely to exist
in most Herbig Ae and T Tauri stars. The concept of such an inner rim
has been widely used to interpret near-IR interferometric data for
Herbig and T Tauri stars (Eisner et al. \cite{E04}, Muzerolle et
al. \cite{M03}, Colavita et al. \cite{CO03}, Eisner et al. \cite{E03},
Monnier and Millan-Gabet \cite{MM02}, 
Millan-Gabet et al. \cite{M01}). Its effects on the disk structure and
emission at larger radii have been  discussed by Dullemond and
Dominik (\cite{DD04}), who propose that the classification of Herbig
Ae stars in two groups, based on the shape of the far-infrared excess
(Meeus et al. \cite{MW01}), can be interpreted as differences between
objects where the outer disk emerges from the shadow of the inner rim
and objects where this does never happen.   
 
In spite of its success in accounting for a variety of observations, 
the actual structure of the rim has not been much discussed. DDN01 
adopted for their models a very crude approximation, namely that the 
illuminated side of the rim is ``vertical", and that its photospheric 
height is controlled by radial heat diffusion behind the rim. Such a 
model, taken at its face value, has the obvious disadvantage that the 
rim emission vanishes for objects seen face-on,  
for which the projection on the line of sight of the rim surface
is null, and for objects seen edge-on, where the rim obscures its own 
emission.  
This is clearly inconsistent with observations of the SED, which show 
that all the Herbig Ae stars  with disks have similar near-IR excess, 
regardless of their inferred inclination (Natta et al.~2001; Dominik
et al. \cite{D03}).   
 
The vertical shape of the illuminated face of the rim is clearly not 
physical, as pointed out already by DDN01. Several effects are likely 
to ``bend" the rim: among them, one can expect that radiation pressure 
on dust grains or dynamical instability, due to self-shadowing effects, 
could modify the illuminated face of the rim (Dullemond 2000;
DDN01). None of these suggestions, however,  has been
explored further.

In this paper, we will  discuss in detail a different
process, not   mentioned so far,
which depends exclusively on the basic physics of
dust evaporation, i.e., on
the dependence of the evaporation temperature on  gas density.
Circumstellar disks are characterized by a very large variation of
the density in the vertical direction, so that
the dust evaporation temperature varies by several hundred degrees in few
scale heights;  moving vertically away from the disk midplane along
the rim, dust will evaporate at lower and lower temperatures, i.e.,
further away from the central star. This very simple effect curves
significantly the inner face of the rim, as we will describe in the
following.  
 
The paper is organized as follows. In \S 2 we describe the model we 
use to compute the rim shape and its observational properties. The 
results are presented in \S 3, where we discuss also how the rim 
depends on dust properties. A discussion of the results, in view
of the existing  observations, follows
in \S 4, and a summary in \S 5.

\section {Model description} 
\label{sec:model} 
 
Our model of the inner rim of passive-irradiated flaring disks joins  
the two different analytical methods to solve the structure of a
circumstellar disk  proposed respectively by Calvet et
al. (\cite{C91}, \cite{C92}, hereafter C92) and Chiang and Goldreich
(\cite{CG97}, hereafter CG97). The temperature in the rim atmosphere
is determined using the analytical solution of the problem of the
radiation transfer as in C92, neglecting the heating term due to the
mass accretion. The vertical structure of the rim is then computed in
a way derived from CG97 and  DDN01, adding a relation between the dust
vaporization temperature and the gas density as proposed in Pollack et  
al. (\cite{PH94}). As a result we obtain a curved model for the inner 
rim whose features are described in the next paragraph. 
  
Although the expressions for the dust temperature derive from a first
order solution of the radiation transfer equation, we found (see
Appendix)  good agreement with the correct numerical result in most
cases. To zero order, we can compute the rim structure avoiding proper
radiation transfer calculations.  
 
In the limit where the incident angle $\alpha$ of the stellar radiation  
onto the disk surface is $\alpha \ll 1 $, the equations of the 
temperature for $\tau_d=0$ and $\tau_d \gg 1$ (where $\tau_d$ is the 
optical depth for the emitted radiation) are formally equal to the 
optically thin $T_s$ and  midplane $T_i$ temperatures introduced  
by CG97 (two-layer approximation). Therefore, while the two layer
approximation is useful to study the structure and the emission
features (e.g. silicate features at $10\mu m$) of the flaring part of
the disk (as in DDN01), it must be abandoned in modeling the inner
rim, since $\alpha \simeq 1$. Nevertheless we can adapt the CG97
treatment of the vertical structure of the disk to the inner rim using
the appropriate expressions for the dust temperature. 
  
In order to clarify this concept and to introduce the relation 
between the vaporization temperature of dust and the gas density, the 
basic equations are briefly summarized. We refer to the cited works for 
a physical discussion of the equations.

 
We suppose that the disk is heated only by the stellar radiation and 
we call $\alpha$ the incident angle beetwen the radiation and the disk  
surface. The incident beam is absorbed exponentially as it penetrates 
the dusts and if $\tau_d$ is the optical depth for the emitted 
radiation, the dust temperature $T(\tau_d)$ is given by the relation 
(C92)  
\begin{eqnarray} 
\label{eq:T_tau} 
  T^4(\tau_d) & = & T^4_\star \left( \frac{R_\star}{2r} \right)^{2} 
  \cdot \nonumber \\ 
  & & \cdot \left[ \mu (2 + 
  3\mu\epsilon) + \left( \frac{1}{\epsilon} - 3\epsilon\mu^2 
  \right)e^{(-\tau_d/\mu\epsilon)} \right]    
\end{eqnarray} 
where $\mu = \sin\alpha$ and $\epsilon$ is an efficiency factor
that characterizes the grain opacity, defined as the ratio of the Planck mean  
opacity of the  grains at the local temperature $T(\tau_d)$ and at the 
stellar temperature $T_\star$, respectively. In writing the previous 
equation we have assumed that the scattering of the dust grain is 
negligible and that the Planck mean opacity is defined as: 
\begin{equation} 
\label{eq:Kp} 
  K_P(T) = \frac {\int_0^\infty B_{\nu}(T)k_{\nu} d\nu}{\int_0^\infty 
  B_{\nu}(T)d\nu}.  
\end{equation} 
Following Muzerolle et al. (\cite{M03}), the continuum emission of the  
disk is assumed to originate from the surface characterized by the 
optical depth $\tau_d=2/3$. 
 
In the flaring part of the disk, for which $\mu \ll 1$, 
Eq.(\ref{eq:T_tau}) can be rewritten in terms of the two 
layer approximation, proposed in CG97, in which the interior of the 
disk (with $\tau_{d} \gg 1$) is heated to the temperature $T_i$ by 
half of the stellar flux, while the other half of the stellar flux is 
reemitted by the superficial layer heated at the optically thin 
temperature $T_s$:  
\begin{equation} 
\label{eq:Ti} 
T^4(\tau_d \gg 1) \simeq  T_i^4 = \frac{\mu }{2} \left( 
\frac{R_\star}{r} \right)^{2} T^4_\star  
\end{equation} 
\begin{equation} 
\label{eq:Ts} 
T^4(\tau_d = 0) \simeq  T_s^4 = \frac{1}{\epsilon} \left( 
\frac{R_\star}{2r} \right)^{2} T^4_\star  
\end{equation} 
Following DDN01 we assume that the dust component of the disk is 
truncated on the inside by dust evaporation, forming an inner 
hole of radius $R_{in}$. Inside this hole, only gas may exist but as 
long as the gas is optically thin to the stellar radiation (Muzerolle 
et al. \cite{MAC04}) we can neglect its absorbing effect.  
Near the inner radius $R_{in}$, since the rim surface is nearly 
perpendicularly exposed to the stellar radiation ($\mu \simeq 1$),  
Eq.~(\ref{eq:Ti}) and (\ref{eq:Ts}) must be replaced by the more
general relations 
\begin{equation} 
\label{eq:Tinf} 
T^4(\tau_d \gg 1)\equiv T^4_\infty = \mu (2 + 3\mu\epsilon) 
\left(\frac{R_\star}{2r} \right)^{2} T^4_\star , 
\end{equation} 
\begin{equation} 
\label{eq:T0} 
T^4(\tau_d = 0) \equiv T^4_0 = \left( 2\mu + \frac{1}{\epsilon}
\right) \left( \frac{R_\star}{2r} \right)^{2} T^4_\star , 
\end{equation} 
derived as limiting cases from Eq.(\ref{eq:T_tau}). 
 
It is useful to note here that, since the dust in the rim is by
definition close to the evaporation temperature, the value of
$\epsilon$ is fixed and depends only on the grain absorbing cross
section. Moreover, to first order, since the difference between $T(\tau_d=0)$
and $T(\tau_d=\infty)$ is never very large, at the inner radius ($\mu
= 1$) the ratio between these two temperature depends only on
$\epsilon$ and is given by:   
\begin{equation} 
\label{eq:T0/Tinf} 
\frac{T_0}{T_\infty}= \left[ \frac{2\epsilon+1}{\epsilon(2+3\epsilon)} 
  \right]^{1/4}.  
\end{equation} 
\begin{figure} 
  \begin{center} 
    \leavevmode 
    \centerline{ \psfig{file=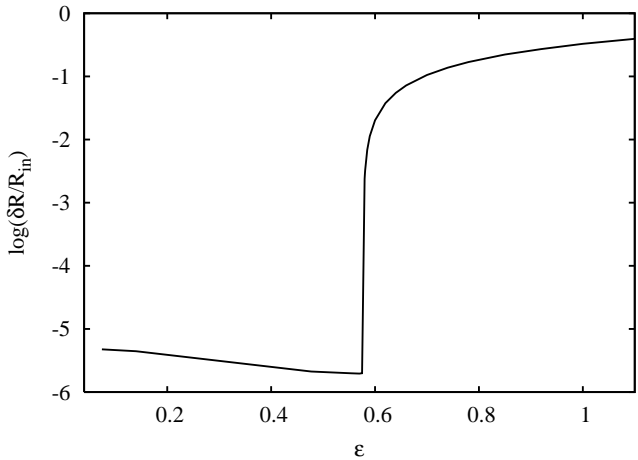,width=9cm,angle=270} } 
  \end{center} 
  \caption{
Width of the region  over which 99\% of the stellar radiation
is absorbed  as function of $\epsilon$.
 } 
  \label{fig:Dr} 
\end{figure} 

The rim  is defined by the condition that the incoming stellar 
radiation is entirely absorbed by the intervening dust and in the
following we will refer to the surface where $\tau_s=1$ to define
its shape and  location. Once the grain evaporation temperature
$T_{evp}$ is fixed, one can see from Eq.(\ref{eq:T_tau}) that the rim
has a sharp surface when $\epsilon < \epsilon_{cr} = 1/\sqrt{3}$. In
this case the optically thin temperature $T_0$ of the grains is higher
than $T_\infty$. As soon as a grain can survive, the optical depth for
the stellar radiation $\tau_s$ increases rapidly from zero to very
high values and the rim surfaces characterized by different values of
the optical depth $\tau_s$ are all compressed in a vary narrow zone,
located at $R_{in}$, i.e., where $T_0 = T_{evp}$. 
For $\epsilon > \epsilon_{cr}$, the dust
temperature increases with the optical depth and the transition to
optically thick regimes is controlled by the geometrical dilution of
the stellar radiation; in this case  the transition from $\tau_s = 0$
to very high values occurs in a relatively broader region
(see Fig. \ref{fig:Dr}); however, also in these cases the rim is located
by the $\tau_s=1$ surface with an accuracy better than 10--20\%.
 
With the assumption that the disk is in hydrostatic equilibrium in the  
gravitational field of the central star and that is isothermal in the 
vertical direction $z$, the gas density distribution is expressed by 
the Gaussian relation 
\begin{equation} 
\label{eq:rho} 
\rho_g(r,z) = \rho_{g,0}(r) \exp(-z^2/2h^2), 
\end{equation} 
where 
\begin{equation} 
\label{eq:h/r} 
\frac{h}{r}= \left( \frac{T_\infty}{T_c} \right)  
             \left( \frac{r}{R_\star } \right) ^{1/2} 
\end{equation}    
defines the relation between the pressure scale $h$ and the interior  
temperature $T_\infty$ at a distance $r$ from the central star.  
The temperature $T_c$ is a measure of the gravitational field of the  
central star, expressed by 
\begin{equation} 
\label{eq:Tc} 
T_c = \frac{G M_\star \mu_g}{k R_\star} 
\end{equation}
where $\mu_g$ is the mean molecular weight of the gas.

Note that the interior temperature of the disk $T_\infty$ depend on
the incident angle $\alpha$ between the stellar radiation and the disk
surface, through  Eq.(\ref{eq:Tinf}). The quantity $\mu = \sin\alpha$
accounts for the projection of a disk annulus  on a
plane perpendicular to the incident radiation and  is given by the
relation  
\begin{equation}
\label{eq:alpha}
\mu = \sin \left( \arctan \left[ \frac{r}{z(r)} \right] + \arctan \left[
  \frac{dz(r)}{dr} \right] - \frac{\pi}{2}  \, \right),
\end{equation}
where $z(r)$ is the axisymmetric equation of the surface of the disk
that we want to determine. For the standard flaring model, the
incident angle $\alpha$ is given by the relation 
\begin{equation}
\label{eq:alphaCG}
\sin\alpha \equiv \mu \simeq \frac{0.4R_\star}{r}+r \frac{d(H/r)}{dr}\, ,
\end{equation}
where $H$ is the photospheric height of the disk, defined as the
height to which the optical depth of the disk to the stellar radiation
is $\tau_s=1$, on a radially directed ray. For the inner rim we can 
retain this definition for $H$ and place $z(r) = H(r)$ in
Eq.(\ref{eq:alpha}), where $H$ is thus given by the relation 
\begin{equation}
\label{eq:Hr}
  \frac{K_P(T_\star)}{\mu} \int_{H(r)}^\infty \rho_d(z,r)dz = 1.
\end{equation} 

As shown in  DDN01, the ratio $\chi = H/h$, between the photospheric 
and pressure height, is a dimensionless number of the order of 4-6,    
depending weakly on the angle $\alpha$, the density $\rho_d$ and the 
Planck mean opacity of dust at the stellar temperature 
$T_\star$. Finally, to obtain the dust density distribution from 
Eq.(\ref{eq:rho}) we assume a constant ratio $\rho_d / \rho_g = 0.01$.   

We can follow CG97 and DDN01 to obtain a self-consistent solution of 
Eq.(\ref{eq:Kp}), (\ref{eq:Tinf}), (\ref{eq:T0}), (\ref{eq:rho}),
(\ref{eq:h/r}), (\ref{eq:alpha}), (\ref{eq:Hr}) to determine the
structure of the disk, as long as dust evaporation can be
neglected. When dust evaporation is important, DDN01 have developed an
approximate solution of the rim/disk structure under the assumption of
constant $T_{evp}$. The inner rim has a vertical surface toward the
star located at the dust evaporation distance $R_{in}$. The vertical
photospherical height $H$ depends on the dust density behind the rim;  
since the rim is higher than the flaring disk, it casts a shadows over 
the disk. In this shadowed  region, assuming that there is no external 
heating except the stellar radiation, the pressure height $h$ depends 
only on the radial heating diffusion. The exact determination of the 
structure of the rim in this  diffusive region would thus require to 
solve the problem of radiation transport in two dimensions (see 
Dullemond \cite{D02}). Since this goes well beyond our aims, we adopt
the approximated relation used in DDN01    
\begin{equation} 
\label{eq:trans} 
\frac{d(rT_\infty^4)}{dr} \simeq -\frac{rT_\infty^4}{h}. 
\end{equation}    
Since $h \propto (T_\infty R^3)^{1/2}$ (from Eq.(\ref{eq:h/r})), the 
temperature $T_\infty$ can be eliminated and for $h \ll R$ we obtain 
the relation for $h$ behind the rim  
\begin{equation} 
\label{eq:dH} 
\frac{d(h/r)}{dr} = - \frac{1}{8R}. 
\end{equation} 
The solution of Eq.(\ref{eq:dH}) can therefore be used to determine 
the density $\rho_d(r,z)$ behind the rim through Eq.(\ref{eq:rho}). 
 
We now introduce the relation between the evaporation temperature of 
the dust and the gas density. The physical reasons for this effect 
can  be easily understood thinking of evaporation as the process
by which equilibrium
between the gas pressure and the surface tension of the dust grains is
reached:
the higher the gas density, the higher will be the evaporation
temperature. We adopt for the dust in the disk the model proposed by
Pollack et al. (\cite{PH94}). In this model, the grains with the
higher evaporation temperature are the silicates, which will therefore
determine the location of the rim. Their evaporation temperature (see
Table 3 in Pollack et al.) varies with the gas density roughly as a
power law, of the kind:  
\begin{equation}  
\label{eq:Tevp} 
T_{evp} = G \rho_g^{\gamma}(r,z_{evp})  
\end{equation} 
where $G = 2000$ and $\gamma =1.95 \cdot 10^{-2}$. 
Since $\rho_g$ varies exponentially with $z$ (see Eq.(\ref{eq:rho})),
for $z_{evp}=h$ the evaporation temperature of dust is only $1\%$
smaller than that on the midplane, but for $z_{evp}= H \sim 5h$ (for a
reasonable value of $\chi$) the difference is about $20\%$ and dust
evaporation takes place at a larger distance from the star, curving
the rim.  
 
To obtain a self-consistent determination of the structure of the 
curved rim, we implemented a numerical method able to solve
Eq.(\ref{eq:Tevp}) together with the set of equations (\ref{eq:Kp}),  
(\ref{eq:Tinf}), (\ref{eq:T0}), (\ref{eq:rho}), (\ref{eq:h/r}), 
(\ref{eq:alpha}), (\ref{eq:Hr}). The distance of dust evaporation in
the midplane $R_{in}$ is computed for $z=0$ in the set of equations
and  is taken as the starting point of the radial grid on which  the
rim structure is computed. As a result of the calculations, we obtain
the location in the $(R,z)$ plane of the rim surfaces characterized by
a constant value of the optical depth. The surface for $\tau_s=0$ is  
thus the evaporation surface of dust grains, for $\tau_s =1$ we 
obtain the surface relative to the photospheric height $H$, defined 
through the Eq.(\ref{eq:Hr}), while for $\tau_d=\epsilon \cdot \tau_s 
= 2/3$ we obtain the emitting surface of the rim. 
 
Moving on the rim surface away from the star, the incident angle 
$\alpha$ decreases and, when it approaches zero, the determination of
the rim shape becomes very difficult. This is mainly due to the fact
that the described solution for the radiation transfer neglects the
heat diffusion between contiguous annulus of the rim. Therefore both
the midplane temperature $T_{\infty}$ and the pressure height $h$ of
the rim goes unphysically to zero for $\alpha = 0$, according to
Eq.(\ref{eq:Tinf}) and (\ref{eq:h/r}). To avoid this unrealistic
behaviour, we use the approximated relation discussed previously (see
Eq.(\ref{eq:dH})) to determine the pressure height $h$ in the region
where the diffusion is the dominant heating source. The transition
distance between the region of the rim heated by the star and those
heated by the diffusion is determined imposing continuity of
$dh/dr$. The photospherical height $H$ is then determined as for the
vertical rim.

\section {Results} 
\label{sec:result} 
 
Using the model described in the previous section, we compute the 
structure of the inner rim for a disk heated by a star with 
temperature $T_{\star}=10000$K, mass $M_{\star} = 2.5 M_{\sun}$  and 
luminosity $L_{\star}=47 L_{\sun}$. We take a disk surface density  
$\Sigma(r)=2\cdot10^3(R/AU)^{-1.5} g\,cm^{-2}$, and 
a dust-to-gas mass ratio $dust/gas = 0.01$. In our models, 
this value of $\Sigma$ corresponds to a midplane gas density 
of about $10^{-8} g/cm^3$ at $0.5$AU from the star. 
Note that the results are not very sensitive to the exact value of
$\Sigma$, as long as the inner disk remains very optically thick. 
 
The dust properties are those of the astronomical silicates of  
Weingartner and Draine (\cite{WD01}).
In our models, we consider that all the grains have 
the same size and characterize their properties with the quantity 
$\epsilon$, the ratio of the mean Planck opacity at the evaporation 
temperature to that at the stellar temperature (see 
Eq.(\ref{eq:Kp})). The evaporation 
temperature of silicate grains vary from ~1600K, for gas 
density of $\rho_g=10^{-6} g/cm^3$, to ~1000K for $\rho_g=10^{-16} 
g/cm^3$ (see Eq.(\ref{eq:Tevp})).   
For the Weingartner silicates and a vaporization 
temperature $T_{evp} \sim 1400$ K, $\epsilon \sim 0.08$ for grains of 
radius $a=0.1 \mu m$, and grows to values of about unity for grain 
radii $>5\mu m$.

\subsection{The Rim Shape} 
\label{sec:shape}
 
\begin{figure*} 
  \begin{center} 
    \leavevmode 
    \centerline{ \psfig{file=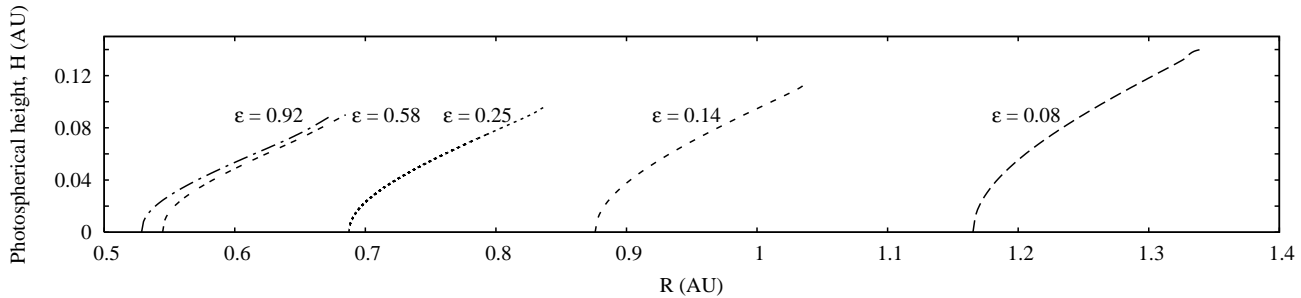,width=18cm,angle=270} } 
  \end{center} 
  \caption{Photospherical height H ($\tau_s=1$) of the inner rim   
for different values of $\epsilon$, as labelled.  
The curves have been computed  for a 
  star with $T_{\star}=10000$K, $M_{\star} = 2.5 M_{\sun}$, 
  $L_{\star}=47 L_{\sun}$. Each curve ends
where the rim becomes optically thin to the stellar radiation. 
  } 
  \label{fig:Zr} 
\end{figure*} 
 
The shape of the rim is shown in Fig. \ref{fig:Zr}, which plots the 
locus of $\tau_s=1$ (i.e., the photospheric height H) as function of R
for different values of $\epsilon$. The ending point of the rim
($R_{out}$, $H_{out}$) is when the rim becomes optically thin at the
stellar radiation.  
 
For $\epsilon \geq \epsilon_{cr} \sim 0.58$, corresponding to silicate
grains bigger than $1.3\mu m$, the inner radius $R_{in}$, the outer
radius $R_{out}$ and the maximum photospheric height of the rim
$H_{out}$, all vary very little, with values  $R_{in} \simeq 0.50$ AU,
$R_{out} \simeq 0.65$ AU and $H_{out} \simeq 0.09$ AU. For smaller
values of $\epsilon$, the  rim gets steeper and the inner radius
increases. For $\epsilon=0.08$, corresponding to grains with  radius
$0.1\mu m$, the rim has $R_{in} = 1.16$ AU, $R_{out} = 1.34$ AU and
$H_{out} = 0.14$ AU.   
 
The shape of the rim can be roughly characterized by the ratio of its  
maximum height $H_{out}$ over width $\Delta R=(R_{out}-R_{in})$. This
quantity, which is nominally infinity in a vertical rim, becomes 
in the curved rim models
$\sim 0.6$ for $\epsilon\simgreat \epsilon_{cr}$ and is $\sim
0.8$ for $\epsilon\sim 0.14-0.08$. In other words, as grains grow, the
inner rim approaches the star but the bending of the surface varies  
very little.   
 
\begin{figure} 
  \begin{center} 
    \leavevmode 
    \centerline{ \psfig{file=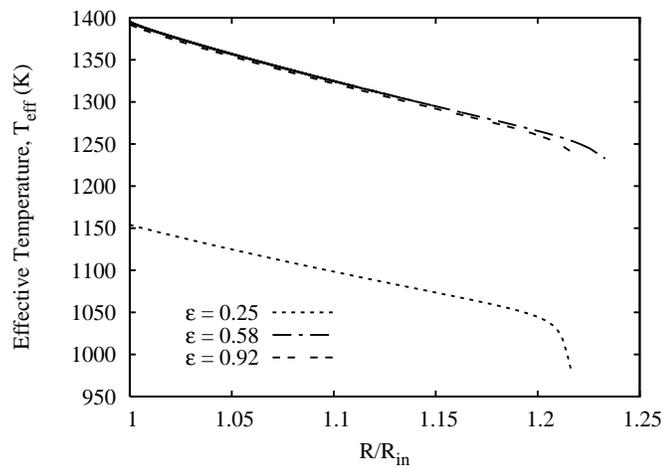,width=9cm,angle=270} } 
  \end{center} 
  \caption{Effective temperature (i.e., $T(\tau_d=2/3)$) along
the rim surface for different values of $\epsilon$, as labelled.
Each curve is plotted as function of $R/R_{in}$, where
  $R_{in}$ is the distance of the rim from the star on the disk midplane.
Note that, in fact, not only $R$ but also $z$ changes along
each curve, as shown in Fig.~\ref{fig:Zr}. For $\epsilon 
  >  \epsilon_{cr} \simeq 0.58$ $T_{eff}$
  is almost independent of the  grain opacity. 
  } 
  \label{fig:Tr} 
\end{figure}

For any given value of $\epsilon$, the dust  temperature along the rim
is not constant (as in the vertical rim) but decreases from values of
about 1400 K, typical of silicate evaporation temperatures  at
density of $10^{-8} g/cm^3$, to 1200 K for density of $10^{-11}
g/cm^3$. Fig. \ref{fig:Tr} plots the effective temperature of the rim
(i.e., $T(\tau_d=2/3)$) along its surface for different values of
$\epsilon$. For $\epsilon \simgreat \epsilon_{cr}$ the effective
temperature is equal to the vaporization temperature; for $\epsilon <
\epsilon_{cr}$ the effective temperature is generally lower since the
dust temperature decreases with the optical depth, according to
Eq.(\ref{eq:T_tau}). This variation, however, is sufficiently small
that most of the rim emission occurs in the near-IR, as for the
vertical rim.    

\subsection {The Rim SED} 
 
\begin{figure} 
  \begin{center} 
    \leavevmode 
    \centerline{ \psfig{file=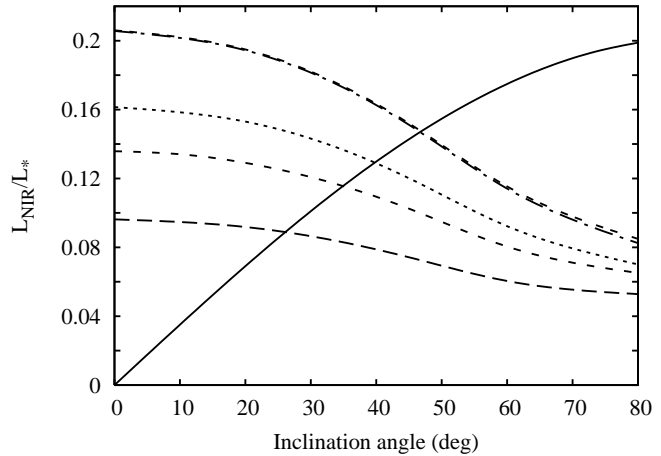,width=9cm,angle=270} } 
  \end{center} 
  \caption{Near infrared emission of the curved rim (integrated
between $1.25 - 7 \mu m$ and normalized to $L_\star$) 
  for different inclination angles and  
  different values of  $\epsilon$. Starting from the 
  bottom, the curves refer to $\epsilon = 0.08$ (long-dashed), 0.14
(short-dashed), 
  0.25 (dotted),  $0.58=\epsilon_{cr}$ (short-dash-dotted),  0.92
(long-dash-dotted). For the last two values of 
  $\epsilon$ the emission is almost the same and the lines 
  overlap. The continuum line plots the emission of the 
  vertical rim, computed for $\epsilon = \epsilon_{cr}$. 
  } 
  \label{fig:fluxes} 
\end{figure}

The fraction of stellar luminosity intercepted by the rim, given by the 
ratio $H_{out}/R_{out}$, varies from $10\%$, for small values of 
$\epsilon$, to $14\%$, for $\epsilon \geq \epsilon_{cr}$.  
Assuming that the rim is in thermal equilibrium, the  intercepted 
radiation is equal to the total emitted flux $F_{IR}$. For each 
$\epsilon$, we have computed the spectral energy distribution of the 
rim emission for different  inclination angles. As discussed in 
\S\ref{sec:model}, the rim emission is computed as that of a blackbody 
at the local temperature along the rim $\tau_d = 2/3$ surface. We will 
come back in the Appendix to this assumption, which, in any case, 
gives a very good approximation to the global properties of the rim 
emission (see also Muzerolle et al. \cite{M03}, C92). Most of the 
emission, as expected from the range of temperatures, occurs in the 
near-IR. We have computed the fraction of the stellar luminosity 
re-emitted by the rim in the wavelength range 1.25-7.0 $\mu$m for 
different values of $\epsilon$ as function of the inclination 
(Fig. \ref{fig:fluxes}). This near-IR excess peaks at zero 
inclination, where has values between $\sim10\%$ (smaller grains) and 
$\sim20\%$ (larger grains). As the inclination increases, the near-IR 
excess decreases slowly, reaching values between 5\% and 8\%, 
depending on $\epsilon$. For inclination higher then $ \sim 80^\circ$ 
the rim emission is self-absorbed. Note that for the large grains, 
with $\epsilon \simgreat \epsilon_{cr}$, the near-IR excess becomes almost 
independent of $\epsilon$.  
 
The behaviour described in Fig. \ref{fig:fluxes} is  very different 
from what one obtains in the case of  the vertical rim. As described 
in the previous section, neglecting the dependence of the evaporation  
temperature of  grains from  gas density results in the inner face of
the rim to be vertical (as in DDN01). With $\epsilon = \epsilon_{cr}$
and $T_{evp} \simeq 1400K$ (equal to the evaporation temperature at
the inner radius of the curved rim), the inner radius $R_{in}$ is the
same as for the curved rim; the photospheric height, evaluated using
the approximation described by Eq.(\ref{eq:dH}), is also the same  
($H_{out} = 0.09$ AU); the fraction of stellar luminosity intercepted 
by the vertical rim is $17\%$. However, the value of $F_{NIR}$ 
observed for different inclination angles is very different, with a 
very strong (and opposite) dependence on 
$i$. (Fig. \ref{fig:fluxes}). In particular, $F_{NIR}$ vanishes for 
face-on disks, and is maximum (of the order of 20\%) for very inclined  
systems, just before self-absorption sets in.  
 
There are also differences between the curved and vertical 
rim models concerning the shape of the predicted 3$\mu$m bump. 
While the vertical rim has a constant temperature  of 1400K over all 
its surface, in the curved rim the temperature varies from 1400K on 
the midplane to about 1200K at the outer edge (see Fig. 
\ref{fig:Tr}) and the SED is broader than  a single-temperature black 
body. In practice, however, this is only  a minor effect.

\subsection {Rim Images} 
\label{sec:vis} 
 
\begin{figure*} 
  \begin{center} 
    \leavevmode
    \centerline{ \psfig{file=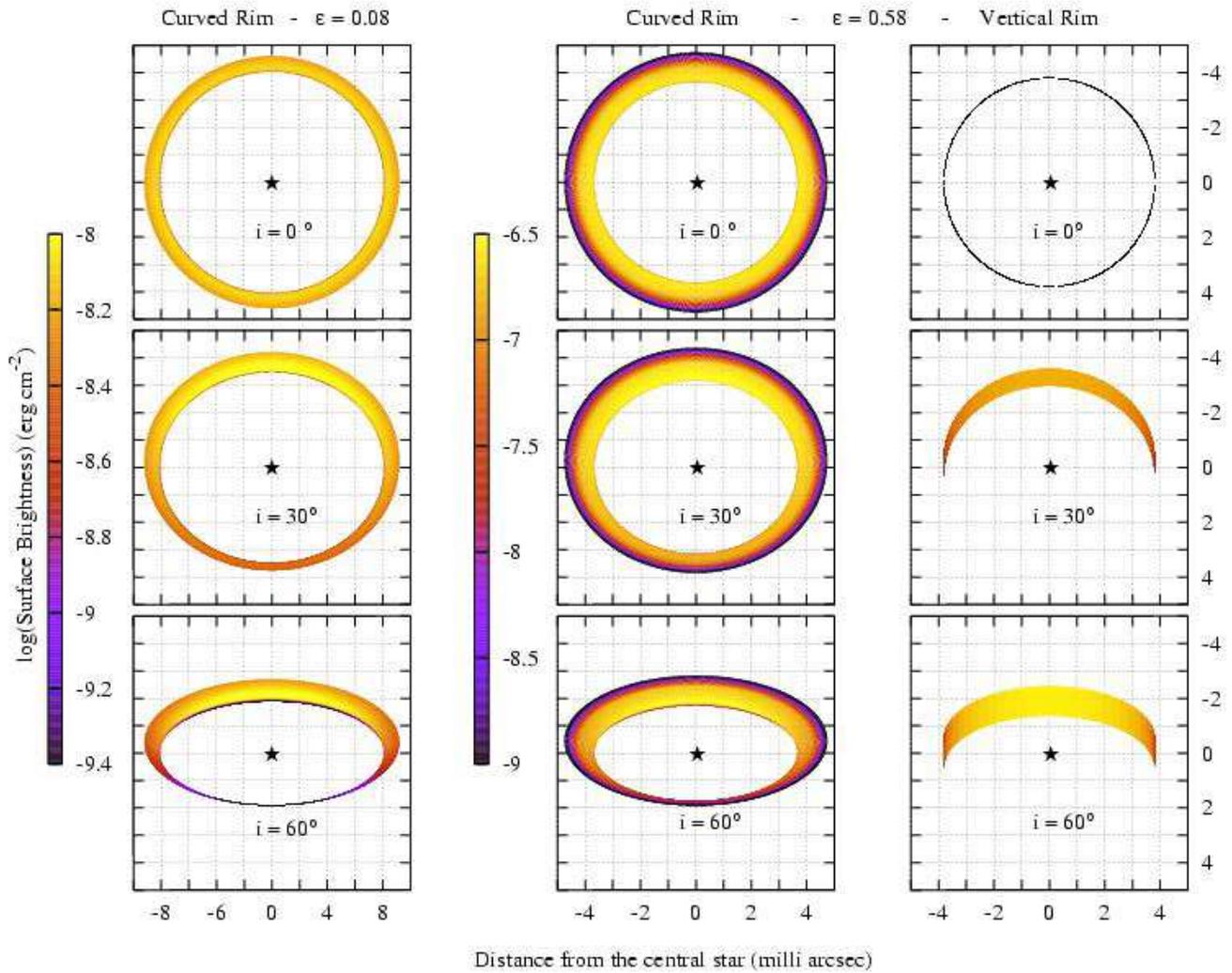,width=20cm,angle=270} } 
  \end{center} 
  \caption{ Synthetic images of the curved rim for different values of
    $\epsilon$ and inclination angle ($i=0^\circ$ is face-on);
    comparison with the vertical rim. The left and central panels show
    the images of the curved rim calculated for $\epsilon = 0.08$
    (silicate grains with radius $a=0.1 \mu m$) and $\epsilon =
    \epsilon_{cr} = 0.58$ ($a=1.3 \mu m$). The right panels show the
    images for the verical rim calculated for $\epsilon = 0.58$.
    The surface brightness of the  rim, plotted  in colors, is
    computed for a wavelength of $2.2 \mu m$. The stellar parameters
    are as in \S 3, the distance is $d=144pc$.
    The  vertical rim for $i=0^{\circ}$  has zero brightness. Note
    that as discussed in \S 3 for $\epsilon>\epsilon_{cr}$
    the images of the curved rim remain the same.
  }
  \label{fig:conf} 
\end{figure*}

Fig. \ref{fig:conf} shows synthetic images of the curved rim at $2.2
\mu$m for grains  with $\epsilon=0.08$ (radius of about 0.1 $\mu$m)
and $\epsilon=\epsilon_{cr}=0.58$ (radius of about 1.3 $\mu$m). As
described in \S 3.1 (see Fig. \ref{fig:Zr} and Fig. \ref{fig:Tr}), the
inner radius of the rim is larger for smaller grains and the surface
brightness is lower, due to the lower effective temperature of the
emitting surface.

For the comparison with the vertical model of the rim, the left panels of
Fig. \ref{fig:conf} show the images of the vertical rim, calculated
for $\epsilon = 0.58$. The largest difference is at low 
inclinations: for $i=0^{\circ}$ (face-on disks) the vertical rim 
vanishes, as the projected emitting surface along the line of sight is
zero, while the curved rim has a centrally symmetric ring shape. 
For a distance of 144pc, the inner radius of the ring, for $\epsilon =
\epsilon_{cr}$, is about 4 
milliarcsecond ($mas$); its brightness  peaks practically at $R_{in}$ 
($R_{peak}/R_{in}\sim 1.05$) and decreases slowly (by about a factor 
of two) outward, until at $R_{out}$ it drops to very low values. At 
this distance, the width of the rim (roughly its FWHM) is about 0.8 
$mas$.   

For higher inclination the central symmetry is lost and the
projected image is an ellipse with one edge brighter than the
other. In general, however, the brightness distribution of the rounded
rim is much more symmetric than that of the vertical rim.

 
\subsection{Grain size distribution} 
 
The results shown so far have been computed assuming that all the 
grains have the same composition and size. This is in practice 
unrealistic, and one wonders how the results will change if  grains
with different properties are  present. Different grains  absorb
differently the stellar radiation and reach very different equilibrium
temperatures. The degree of complexity of the radiation transfer
problem increases remarkably (see Wolfire and Cassinelli \cite{WC86}, 
\cite{WC87}) 
and there is no approximate solution such
as Eq.(\ref{eq:T_tau}). In fact, to the best of our knowledge even
{\it numerical} solutions are not available to describe accurately the
transition region where some grains are cooler than evaporation and
survive while others do not.  

We can however estimate the effect of a grain mixture in the following
way. Let us consider for simplicity a classical MRN grain size
distribution, characterized by a power law $a^{-q}$ with $a$ the
radius of the dust grains, $q$=3.5, and $a$ varying between a minimum
value $a_{min}=0.01$ $\mu$m and a maximum value $a_{max} \gg
a_{min}$. From what we have discussed in \textsection\ref{sec:model},
if $a_{max}$ is smaller than $1.3 \mu m$ (corresponding to $\epsilon =
\epsilon_{cr}$ for the adopted silicate grains), all the  grains have
a maximum temperature  at  $\tau_d =0$ and the higher is the value of
$\epsilon$, the smaller is the evaporation distance from the star. As
soon as the largest grains can survive, the stellar radiation is
rapidly absorbed and all the other grains will also survive. In this
case,  the shape of the inner rim is controlled  by the largest grains
in the distribution. The shapes in Fig. \ref{fig:Zr} and the emitted
fluxes shown in Fig. \ref{fig:fluxes}  should not change significantly
as long as one interprets $\epsilon$ as relative to the largest
grains.  
 
The situation is more complex if $a_{max}> 1.3 \mu m$, since the
grains with $\epsilon > \epsilon_{cr}$ can survive near  the star only
in an optically thin regime. However, our single-grain models show
that there is very little difference in the location and shape of the
rim as soon as $\epsilon >\epsilon_{cr}$. This suggests that varying
$a_{max}$ above the ``transition" value (i.e., the value of $\epsilon$
at which the dependence of $T$ on $\tau$ changes from decreasing to
increasing) will not change the rim properties any further, at least
to zero order.  

It is also likely that  grains have not just different size
but also different chemical composition. In this case, the rim
location and properties are determined by the dust species with the
highest evaporation temperature, as long as its contribution to the
disk opacity is sufficient to make the disk optically thick. As
discussed in \S 2, in the Pollack et al.~(1994) dust model silicates
have the highest evaporation temperature. This is why we have
performed all our calculations for silicate grains. However, one
should keep in mind that, in some dust models, graphite contributes
most of the opacity at short wavelengths; since graphite has a much
higher evaporation temperature ($T_{evp} \sim$ 2000 K) than silicates,
its presence would move the rim  much closer to the central
star. Although the details of the rim shape may change, its curving,
which is caused only by the dependence of $T_{evp}$ on the density, 
will not disappear.

\section{Discussion}  
\label{sec:discussion} 
 
\subsection{The 3$\mu$m bump} 
 
\begin{figure} 
  \begin{center} 
    \leavevmode 
    \centerline{ \psfig{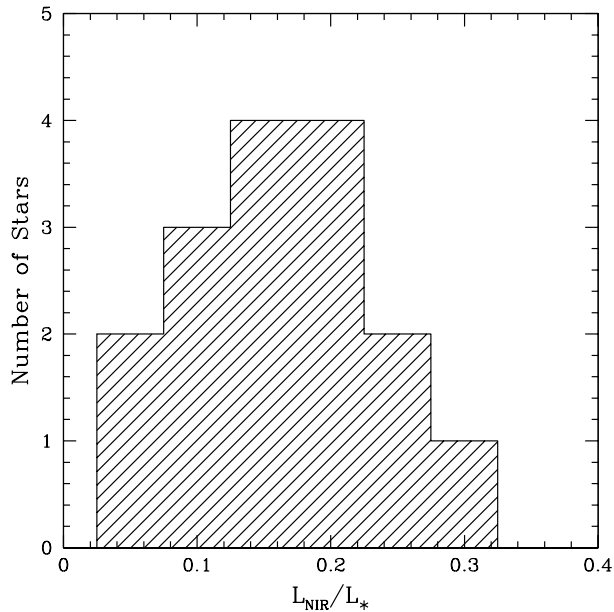} } 
  \end{center} 
  \caption{Histogram of the observed values of $L_{NIR}/L_\star$ for a
    sample of 16 Herbig Ae stars from Natta et al. (2001) and Dominik et
    al. (2003). Note that the Dominik et al. values are computed between
    2 and 7$\mu$m, while those from Natta et al. between 1.25 and 7
    $\mu$m. The uncertainties are in most cases rather large, due to
    the subtraction of the photospheric flux, which contributes
    significantly to the observed ones at short wavelengths, and to
    variability (see Natta et al. for some examples).  
  } 
  \label{fig:nir_hist} 
\end{figure}

The curvature of the inner side of the rim has important consequences
on the observable near-IR excess, as shown in Fig.~\ref{fig:fluxes},
which  shows that the predicted near-IR excess ranges between about 10
and 20\%, and that there is no significant dependence on the
inclination of the disk with respect to the observer. These results
compare very well with the existing observations.
Fig.~\ref{fig:nir_hist} shows the observed values for a well-studied
sample of Herbig Ae stars from Natta et al. (2001) and Dominik et
al. (2003); of a total of 16 objects,  only one (HD~142527) has
$L_{NIR}/L_\star > 0.25$, and one (HD~169142) $<0.09$.   

The observations of the near-IR SEDs of Herbig Ae stars have also 
shown that there is no systematic variation of the near-IR excess with 
the inclination of the disk. The sample of Fig.~\ref{fig:nir_hist} 
includes some rather face-on objects such as AB~Aur ($L_{NIR}/L_\star  
\sim 0.20$, $i \sim20^{\circ}-40^{\circ}$; Fukagawa et al. \cite{F04}) 
and HD~163296 ($L_{NIR}/L_\star\sim 0.21$, $i \sim  30^{\circ}$; 
Grady et al. \cite{G00})  and some UXORS variable, such as UX~Ori,
WW~Vul and CQ~Tau, which are generally considered to be close to
edge-on, and have values of $L_{NIR}/L_\star \sim 0.12-0.25$.  This,
which has been a puzzle for the vertical rim, finds a natural physical
explanation in our models, which predict that the near-IR excess
depends little on the inclination  and that, in particular, does not
vanish for face-on objects. Actually, our models predict that the
largest excesses should be seen in face-on objects, and it would be  
interesting to explore in detail if this is indeed the case. However, 
for such a study to be significant, one would need a large sample of 
objects with known inclinations, which is at present not available. 
 
Another interesting aspect of our results is the dependence of 
$L_{NIR}/L_\star$ on grain properties. If taken at face-value, one 
should expect that only objects with relatively large grains (greater 
than about 1 $\mu$m) can have large values of $L_{NIR}/L_\star$, and
that low values of the near-IR excess can only occur in face-on
objects with small grains. These properties of the rim emission  are
potentially important for a better understanding of grain properties
in the inner disk and should be pursued further in the future,
combining observations of  disks with well known inclinations with
models that explore a broader range of grain properties than we
considered in this paper.

\subsection{The rim radius} 
\label{sec:interf} 
 
\begin{figure}
  \begin{center}
    \leavevmode
    \centerline{ \psfig{file=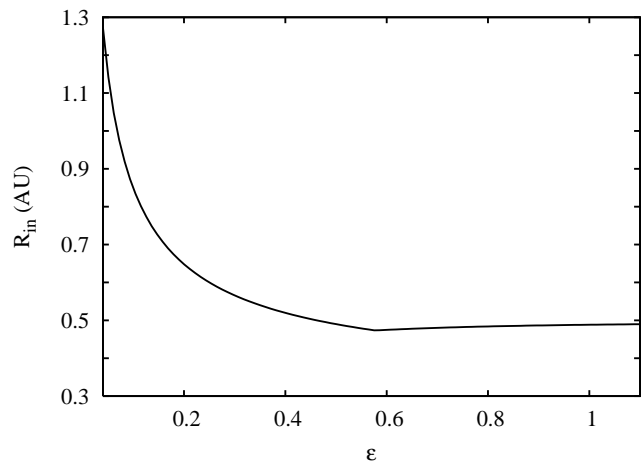,width=9cm,angle=270} }
  \end{center}
  \caption{Behaviour of the inner radius of the rim for
  different values of $\epsilon$, evaluated for the
  star and disk parameters described in \S\ref{sec:result}. Note that
  for $\epsilon \simgreat \epsilon_{cr} \sim 0.58$ (corresponding to silicate
  grains of $1.3\mu$m) the radius of the inner rim is almost constant
  around 0.5AU.  
}
  \label{fig:Rin}
\end{figure}

One side product of our models is  $R_{in}$, the distance from the star to the
rim on the disk midplane. For a face-on disk, $R_{in}$ practically coincide with the position of the peak  of the near-IR brightness. 
For  fixed values of the stellar parameters, $R_{in}$ depends on the dust properties; its value
is shown in Fig.~\ref{fig:Rin}
for different values of $\epsilon$. Larger grains
can live closer to the star than smaller grains and, because of the
inversion in the temperature gradient, $R_{in}$ does not change much
with the grain size  once this exceeds the value for which
$\epsilon=\epsilon_{cr}$, so that for $L_\star=47 L_\odot$,  
$R_{in}\simgreat  0.5$AU. $R_{in}$ scales roughly as $L_\star^{0.5}$,
so that one can expect $R_{in} \simgreat 0.1$ AU for  $L_\star \sim 2 L_\odot$.
These values, especially for small grain sizes, are larger 
than the predictions of simple, optically thin calculations, because 
Eq.(1) correctly includes the effect of the diffuse radiation 
field. Note that if backward scattering cannot be neglected, the 
temperature at $\tau=0$ will be even higher, increasing further the 
size of the inner hole (see Appendix).

Values of $R_{in}$ have been derived for a number of objects from
near-IR interferometric observations
( see, for example, Akeson et al. \cite{A00},
\cite{A02}, Millan-Gabet et al. \cite{M01}, Monnier and
Millan-Gabet \cite{MM02}, Colavita et al. \cite{CO03}, Eisner et 
al. \cite{E03}, \cite{E04}, \cite{E05}). The results  depend
not only on the assumed disk model, but also on other quantities such as the
stellar SED, the disk inclination etc., so that
a direct comparison with our  values of $R_{in}$
can be misleading. We note, however, that  $R_{in} \sim 0.5$ AU
for intermediate-mass objects is roughly consistent with the observations,
suggesting that in many objects grains have grown to sizes
$a \simgreat 1.3 \mu$m (see also Monnier and Millan-Gabet \cite{MM02}). 

This conclusion, of course, needs to be taken with great caution.
There are a number of effects that can change the model-predicted $R_{in}$,
in addition to different grain sizes.
For example, on one side smaller values of $R_{in}$ can be due to
some
low-density gas in the inner disk hole, able to absorb the UV
continuum from the star 
(Monnier and Millan-Gabet \cite{MM02}, Akeson at al. \cite{A05}).
Also,  the presence of same graphite,
which has an evaporation temperature of about 2000 K, can
decrease $R_{in}$.
On the other side, the presence of accretion at a significant rate can
increase $R_{in}$ by heating grains in the inner disk to temperatures higher
than those produced by the photospheric radiation alone (Muzerolle et al. 
\cite{MAC04}).

In practice, one needs to compare in detail
model predictions of the quantities measured  with the interferometers
(visibility curves, their
dependence on baseline and hour angle, phase measurements) for
specific, well known objects. We are currently performing such a study 
with the aim of assessing the impact of curved rim models on the
interpretation of current (and future) interferometric observations.
In particular, we will include a discussion of the effect of the
asymmetries of the rim projected images when not face-on.
The results will be presented in a forthcoming paper (Isella et al. 2005).


\subsection{LkHa101}

To the best of our knowledge, the only image of the inner region of a
circumstellar disk is that obtained by Tuthill et al. (\cite{T01}),
for LkHa101 using Keck in the H and K bands. The results are
reminescent of our images, showing an elliptical ring with a side much
brighter than the other. The similarity is very interesting but one
should keep in mind that LkHa101 is an early B star with a
luminosity between 500 and 50000 \Lsun, depending on its distance. 
Our model may not apply to such an object (e.g., Monnier et al. \cite{MM05}). 

\section {Summary and Conclusions}

In this paper, we discuss the properties  of the inner puffed-up rim
which forms in circumstellar disks when dust evaporates. The rim
existence has been argued for, starting from the work of Natta et
al. (2001), both on theoretical and observational grounds. Here, we
investigate the shape of the illuminated face of the rim. We argue
that this shape is controlled by a fundamental property of
circumstellar disks, namely their very large vertical density
gradient, through the dependence of grain evaporation temperature on
density. As a result, the bright side of the rim is naturally {\it
  curved}, rather than {\it vertical}, as expected when a constant
evaporation temperature is assumed. 

We have computed a number of rim models, which take into account this
effect in a self-consistent way. A number of approximations have been
necessary to perform the calculations, and we discuss their
validity. We think that the basic result (i.e., the curved shape of the rim
illuminated face) is in fact quite robust.

For a given star, the rim properties depend mostly on the properties
of the grains, and very little on those of the disk itself, for
example the exact value of the surface density.  The distance of the
rim from the star is determined by the evaporation temperature (at the
density of the disk midplane) of the dust species that has the highest
evaporation temperature, as long as its opacity is sufficient to make
the disk very optically thick; in the  model of Pollack et
al. (\cite{PH94}) of the dust in accretion disks, silicates have the highest
evaporation temperature. Therefore, we have assumed in our models dust
made of astronomical silicates, and varied  their size over a large
range of values. We find that the rim properties do not depend on
size as soon as $a \simgreat 1.3\mu$m; the values of the rim radii
observed with interferometers suggest that in many pre-main--sequence
disks grains have grown to sizes of 1--few $\mu$m at least. 

The curved rim (as the vertical rim) emits most of its radiation in
the near and mid-IR, and provides a simple explanation to the observed
values of the near-IR excess (the ``3 $\mu$m bump" of  Herbig Ae
stars). Contrary to the vertical rim, for curved rims the near-IR
excess does not depend much on the inclination, being maximum for
face-on objects and only somewhat smaller for highly inclined
ones. This is in agreement with the apparent similarity of the
observed near-IR SED between objects seen face-on and close to
edge-on.  

Finally, we have computed synthetic images of the curved rim seen
under different inclinations. Face-on rims are seen as bright,
centrally symmetric rings on the sky; increasing the inclination, the
rim takes an elliptical shape, with one side brighter than the
other. However,  the brightness distribution of curved rims remains at any
inclination  much more centrally symmetric
than that of vertical ones. 
In a forthcoming paper (Isella et al.~2005) we will discuss the application
of the curved rim models to the interpretation of near-IR interferometric
observations of disks.

\begin{acknowledgements}
We would like to thank Endrik Kr\"ugel for having allowed us to make use of his
radiation transfer code, Leonardo Testi, Kees Dullemond, Carsten Dominik, 
Giuseppe Lodato and Giuseppe Bertin for useful discussions. 
The authors acknowledge partial support by MIUR COFIN grant 2003/027003-001. 
\end{acknowledgements}

\begin{appendix} 
\section {Appendix} 
In this appendix we discuss in detail some of the assumptions on
radiation transfer made in building our rim models. We use as
templates the results of a radiation transfer code developed by
E. Kr\"ugel, which is described in Habart et al. (2004). The code
considers a plane-parallel slab of dust, illuminated on one side by a
star with properties as in \S 3.  
 
\subsection {Evaluation of Eq.(1) for single grains} 
 
\begin{figure} 
  \begin{center} 
    \leavevmode 
    \centerline{ \psfig{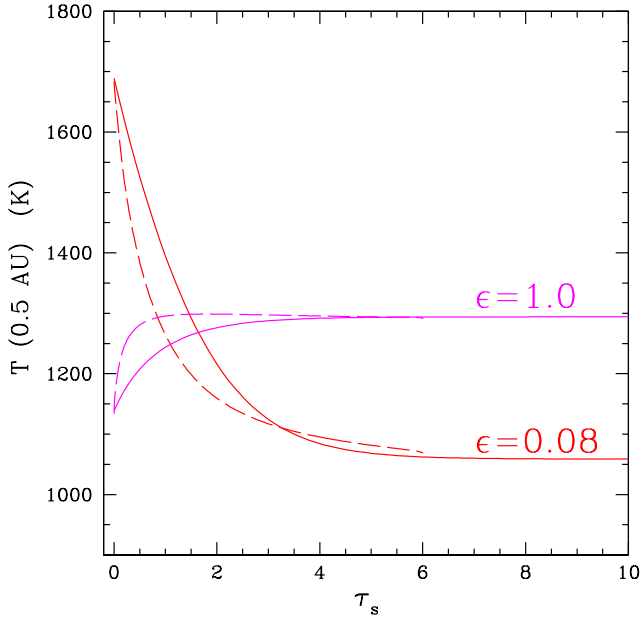} } 
  \end{center} 
  \caption{Temperature as function of the optical depth to the stellar 
    radiation $\tau_s$ for small grains ($\epsilon=0.08$) and very 
    large grains ($\epsilon=1$). The dashed curves are the results of 
    the radiation transfer code, the solid lines the temperature 
    obtained from Eq.(\ref{eq:T_tau}). 
    } 
  \label{fig:app_1} 
\end{figure}

We show in Fig.~\ref{fig:app_1} the comparison of the temperature
derived from Eq.(1) with the results of the radiation transfer code
for two cases, one with $\epsilon=0.08$ (small grains) and one with
$\epsilon=1$ (very large grains). In both cases scattering is
neglected. The temperature is plotted as function of the optical depth
to the stellar radiation $\tau_s$. One can see that Eq.(1) gives the
correct values of $T$ for $\tau_s=0$ and for $\tau_s = \infty$. For
intermediate values of $\tau_s$, the results are in both cases
accurate within 10\%. This is more than adequate for the purposes of
this paper.   

\subsection {SED} 
We compare now the SED of a black-body at temperature $T(\tau_d=2/3)$
from Eq.(\ref{eq:T_tau}), adopted in this paper to describe the rim
emission (see \S 3.2) to the results of the radiation transfer code
(dashed curves)  for the same grains of Fig. \ref{fig:app_1}. The
approximation is very good for large grains. For small values of
$\epsilon$, the discrepancy is larger, and, in particular, one cannot
reproduce any dust feature. However, also in the case $\epsilon=0.08$,
the difference in $L_{NIR}/L_\star$ is only of 4\%.   
 
\begin{figure} 
  \begin{center} 
    \leavevmode 
    \centerline{ \psfig{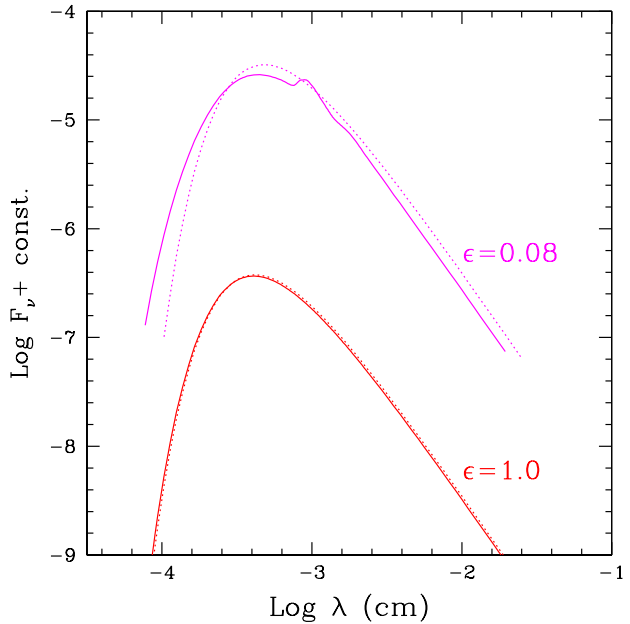} } 
  \end{center} 
  \caption{Comparison of the SED computed with the radiation transfer
    code (dotted lines) and as a black-body at $T(\tau_d=2/3)$ from
    Eq.(1) (solid line) for $\epsilon=0.08$ and $\epsilon=1.0$, as
    labelled. }  
  \label{fig:app_2} 
\end{figure} 
 
\subsection {Scattering} 
\begin{figure} 
  \begin{center} 
    \leavevmode 
    \centerline{ \psfig{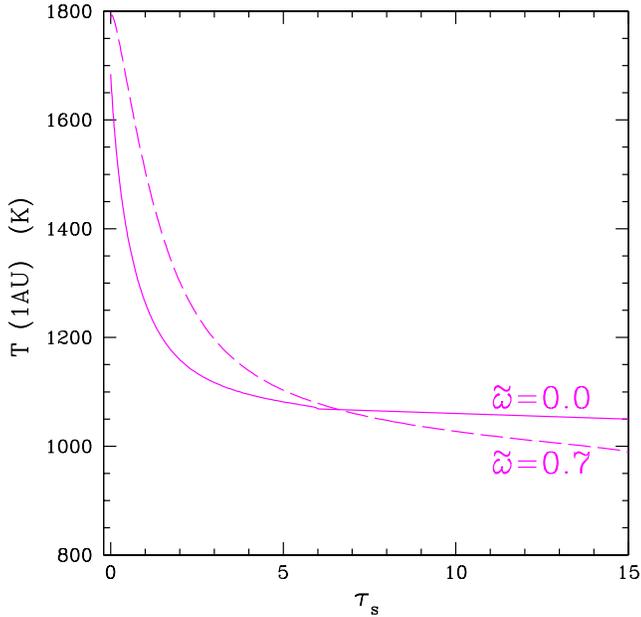} } 
  \end{center} 
  \caption{Temperature profile for grains with $\epsilon=0.08$, 
  when scattering is included ($\tilde{\omega}=0.7$; dashed line) 
or suppressed ($\tilde{\omega}=0.0$; solid line). 
  }
  \label{fig:app_3} 
\end{figure}

Scattering has the obvious effect of increasing the grain temperature  
at the slab surface and decreasing it at large optical depth. We
compare in Fig.~\ref{fig:app_3} the temperature profile of grains with
$\epsilon=0.08$ when scattering is properly included in the radiation
transfer (dashed line) rather than suppressed ($Q_{sc}=0.$). The
curved are labelled with the values of $\tilde{\omega}$, defined as
${Q_{sc}(1-g)/Q_{sc}(1-g)+Q_{abs}}$, where $g$ measures the asymmetry
of the scattering phase function; $\tilde{\omega}=0$ if $Q_{abs}=0$ or
if the scattering is forward peaked ($g=1$). All quantities are
averaged over the stellar radiation field and the corresponding SEDs
show that the value of $L_{NIR}/L_\star$ decreases when scattering is
included by about 25\%. 

\end{appendix}


\begin{thebibliography}{} 
 
 
\bibitem[2000]{A00}
Akeson R.L., Ciardi D.R., van Belle G.T., Creech-Eakman M.J., Lada
E.A. 2000, ApJ 543, 313

\bibitem[2002]{A02}
Akeson R.L., Ciardi D.R., van Belle G.T., Creech-Eakman M.J., 2000,
ApJ 566, 1124 

\bibitem[2005]{A05}
Akeson R.L, Walker C.H., Wood K. et al. 2005, ApJ in press

\bibitem[1991]{C91} 
Calvet N., Pati\~no A., Magris G.C., D'Alessio P. 1991, ApJ 380, 617C  
 
\bibitem[1992]{C92} 
Calvet N., Magris G.C., Pati\~no A., D'Alessio P. 1992, 
Rev. Mexicana Astron. Astrof., 24, 27 
 
\bibitem[1997]{CG97} 
Chiang E.I., Goldreich P. 1997, ApJ 490, 368 
 
\bibitem[2001]{Ch01} 
Chiang E.I., Joung M.K., Creech-Eakman M.J., Qi C., Kessler J.E., 
Blake G.A., van Dishoeck E.F. 2001, ApJ, 547, 1077 
 
\bibitem[2003]{CO03} 
Colavita M. et al, 2003, ApJ, 592, 83  
 
\bibitem[1985]{D85} 
Draine B.T. 1985, ApJS 57, 587D 
 
\bibitem[2003]{D03} 
Dominik C., Dullemond C.P., Waters L.B.F.M., Walch S. 2003, A\&A, 398,
607  
 
\bibitem[2000]{D00} 
Dullemond C.P. 2000, A\&A, 361, L17 
 
\bibitem[2001]{DDN01} 
Dullemond C.P., Dominik C., Natta A. 2001, ApJ 560, 957 
 
\bibitem[2002]{D02} 
Dullemond C.P. 2002, A\&A, 395, 853 

\bibitem[2004]{DD04} 
Dullemond C.P., Dominik C. 2004, A\&A, 417, 159 

\bibitem[2003]{E03}
Eisner J.A., Lane B.F., Akeson R.L., Hillebrand L.A., Sargent
A.I. 2003, ApJ 588, 360

\bibitem[2004]{E04} 
Eisner J.A., Lane B.F., Hillebrand L.A., Akeson R.L., Sargent
A.I. 2004, ApJ 613, 1049 

\bibitem[2005]{E05}
Eisner J.A., Hillebrand L.A., White R.J., Akeson R.L., Sargent
A.I. 2005, astro-ph/0501308

\bibitem[2004]{F04} 
Fukagawa et al., ApJ, 605, 53  

\bibitem[2000]{G00} 
Grady C.A. 2000, ApJ, 544, 859 

\bibitem[2004]{H04}
Habart E., Natta A., Kr\"ugel E. 2004, A\&A, 427, 179

\bibitem[2005]{I05}
Isella A., Testi L., Natta A. 2005, in prep.

\bibitem[2001]{MW01} 
Meeus G., Waters L.B.F.M., Bouwman J., van den Ancker M.E., Waelkens 
C., Malfait K. 2001, A\&A, 365, 476 
 
\bibitem[2001]{M01} 
Millan-Gabet R., Schloerb P. F., Traub W. A. 2001, ApJ, 546, 358 
 
\bibitem[2002]{MM02} 
Monnier J. D., Millan-Gabet R. 2002, ApJ 597, 694 

\bibitem[2005]{MM05}
Monnier J.D., Millan-Gabet R., Billmeier R. et al. 2005, ApJ in press
 
\bibitem[2003]{M03} 
Muzerolle J., Calvet N., Hartmann L., D'Alessio P. 2003, ApJ 597, 149 
 
\bibitem[2004]{MAC04} 
Muzerolle J., D'Alessio P., Calvet N., Hartmann L. 2004, ApJ, 617, 406 
 
\bibitem[2001]{N01} 
Natta A., Prusti T., Neri R., Grinin V. P., Mannings V. 2001, A\&A, 
371, 186 
 
\bibitem[2002]{NRC++} 
Press W. H., Teukolsky S. A., Vetterling W. T., Flannery B. P. 2002, 
Numerical Recipes in C++, Cambrige University Press  
 
\bibitem[1994]{PH94}  
Pollack J.B., Hollenbach D., Beckwith S., Simonelli D.P., Roush T.,
Fong W. 1994, ApJ 421, 615 

\bibitem[2001]{T01}
Tuthill P.G., Monnier J.D., Danchi W.C., 2001, Nature, 409  

\bibitem[2001]{WD01}
Weingartner J.C., Draine B.T. 2001, ApJ, 548, 296 

\bibitem[1986]{WC86}
Wolfire M.G., Cassinelli J.P., 1986, ApJ, 310, 207

\bibitem[1987]{WC87}
Wolfire M.G., Cassinelli J.P., 1987, ApJ, 319, 850





\end{thebibliography}
\end{document}